\documentclass[twocolumn,showpacs,preprintnumbers,amsmath,amssymb]{revtex4}

\usepackage{graphicx}
\usepackage{dcolumn}
\usepackage{bm}
\usepackage{latexsym}
\usepackage{amsfonts}
\usepackage{amssymb}
\usepackage{amsmath}

\begin{document}

\preprint{KUNS-1911}

\title{ Quasi-Thick Codimension 2  Braneworld }

\author{Sugumi Kanno}
\email{sugumi@tap.scphys.kyoto-u.ac.jp}
\author{Jiro Soda}
\email{jiro@tap.scphys.kyoto-u.ac.jp}
\affiliation{
 Department of Physics,  Kyoto University, Kyoto 606-8501, Japan
}%

\date{\today}

\begin{abstract}
We study a codimension 2  braneworld
 in the Einstein Gauss-Bonnet gravity.  
 We carefully examine the structure of possible singularities
 in the system which characterize the braneworld through matching conditions. 
 Consequently, we find that the 
 thickness of the brane can be incorporated as the 
 distributional source, which we dub quasi-thickness.
 On the basis of our formalism, we analyze the linearized gravity 
  and show the conventional Einstein gravity can be recovered on the brane.
 In the nonlinear regime, however,  we find corrections due to the thickness 
 and the bulk geometry. 
 We also point out a possibility that the thickness plays
 a role of the dark energy/dark matter in the universe.  
\end{abstract}

\maketitle


\section{Introduction}

 The old idea~\cite{Rubakov} 
 that our universe may be a braneworld embedded in a 
 higher dimensional spacetime is renewed by the recent development
 in string theory which can be  consistently formulated only in 10 
 dimensions~\cite{Polchinski}. 
 Instead of 10 dimensions, however, 
 most studies of braneworld cosmology have been 
 devoted to 5-dimensional models~\cite{RS1}. 
 This is due to the difficulty in treating 
 the higher codimension object in a relativistic
 manner~\cite{Geroch}.  Nevertheless, it is important to explore the
 possibility of  the higher codimension braneworld~\cite{codimension2}.

 It is the Einstein Gauss-Bonnet gravity that is a natural framework in 
 6-dimensions. Because a natural extension of Einstein gravity to higher 
 dimensions is  the Lovelock gravity and it reduces to
 the Einstein Gauss-Bonnet gravity in 6-dimensions~\cite{Deruelle}. 
 Moreover, the Gauss-Bonnet term appears in the low energy effective action
 in string theory. 
 Therefore, it seems  reasonable to construct a codimension 2 
 braneworld  in the Einstein Gauss-Bonnet gravity. 
  In fact, Bostock et al. claimed 
  that 6-dimensional Einstein Gauss-Bonnet gravity leads to a thin 
  braneworld where a conventional Einstein gravity holds~\cite{Bostock}. 
 More interestingly, they  suggested that  the deviation to the conventional 
 Einstein theory can be possible when the thickness of the brane 
 is taken into account and consequently the variation of the deficit angle 
  is allowed. However, they have never given a precise scheme to
 calculate these corrections.  Rather, they commented that the corrections 
 can not be fixed exactly by the bulk equations. 
 Therefore, it is important to construct a viable 
 concrete model  for the thick braneworld and clarify if corrections
 can be obtainable or not. 
   
  In this paper, we investigate a codimension 2 braneworld 
  in the Einstein Gauss-Bonnet gravity taking into account the thickness. 
  It is difficult to treat the finite thickness as it is.  However,  
  by examining the structure of the singularity in the equations of motion,
  we find a possibility to treat the thickness within the context of the
  distributional source. We name it a quasi-thick
  braneworld. This is the important point in our work which gives
  a framework to clarify various issues in the codimension 2 braneworld.
    In the case of the linearized gravity, we clarify the effect of the 
  bulk on the brane  by solving the whole set of equations of motion in the 
  bulk with proper considerations of the boundary conditions.   
  It turns out that 
  the conventional Einstein gravity is recovered at the linear level.
  However, we also show some corrections due to the thickness can be
  expected at the second order level. The effect of the bulk geometry
  in the nonlinear regime is also discussed. 
  We conclude that corrections to the conventional Einstein
  theory can be fully obtainable  in the case of the quasi-thick
  braneworld. We also point out an interesting possibility 
  that the thickness plays
 a role of the dark energy/dark matter in the universe. 

The organization of this paper is as follows.
 In sec.II, we present a model for thick braneworld
 and derive necessary equations.
 The structure of the singularity is carefully examined. 
 Then, we introduce an idea of the thick braneworld within
 the context of the distributional source, which we dub quasi-thickness.
 In sec.III, we set the background vacuum spacetime. 
 In sec.IV, we solve the bulk geometry using linear perturbation theory.
 In sec.V, the effective theory for the codimension 2 braneworld is 
 presented both in the linear and nonlinear regime. 
  The consistency of our formalism is emphasized here. 
 In the final section, we summarize our results and discuss
  possible applications and extensions. 
 
\section{Quasi-thick Braneworld: A Model for Thick Braneworld}

We consider a codimension 2 braneworld  with a positive tension in 
the 6-dimensional bulk spacetime. 
The Lovelock gravity is a natural framework in higher dimensions and
reduces to the Einstein  Gauss-Bonnet gravity in 6-dimensions. 
 Moreover, the Gauss-Bonnet term ubiquitously appears in the low energy limit
 of string theory. 
  For this reason, the Gauss-Bonnet (GB) term is 
introduced in our model which is described by the action
\begin{eqnarray}
S&=&\frac{1}{2\kappa^2}\int d^6x\sqrt{-g_{(6)}}
	\left[R + \alpha R^2_{\rm GB}\right]
	-\int d^4x\sqrt{-g}~\sigma\nonumber\\
&&\hspace{5mm}
	+\int d^4x\sqrt{-g}{\cal L}_{\rm matter}  \ ,
\end{eqnarray}
where $\kappa^2$ is the 6-dimensional gravitational constant, 
$g_{(6)\mu\nu}$ and $g_{\mu\nu}$ are the 6-dimensional bulk 
and the our 4-dimensional brane metrics, respectively. 
Here, ${\cal L}_{\rm matter}$ is the Lagrangian density of the matter 
on the brane, and $\sigma$ is the brane tension. 
The GB term is given by
\begin{eqnarray}
R^2_{\rm GB}=R^{abcd}R_{abcd}
	-4R^{ab}R_{ab}+R^2  \ .
	\label{GBterm}
\end{eqnarray}
The Latin indices $\{a,b,\cdots\}$ and the Greek indices $\{\mu,\nu,\cdots\}$
are used for tensors defined in the bulk and on the brane, respectively.
The 6-dimensional Einstein equation derived by varying the
above action with respect to $g_{(6)}^{ab}$ takes the form
\begin{eqnarray}
G_{ab}+\alpha H_{ab}=\kappa^2T_{ab} \ ,
\label{EGB}
\end{eqnarray}
where $\alpha$ is the GB coupling constant with dimension
$[\alpha]=L^2$ and 
\begin{eqnarray}
H_{ab}&=&-\frac{1}{2}g_{ab} R_{GB}^2 + 2 R R_{ab} 
	- 4 R_{ad} R^d{}_b \nonumber\\
&&	-4 R^{de} R_{adbe} + 2 R_a{}^{def} R_{bdef}  \ .
\end{eqnarray}
is an analogue of the Einstein
tensor stemmed from the GB term (\ref{GBterm}).

 We will assume that a 6-dimensional metric has axial symmetry, which reads 
\begin{eqnarray}
ds^2=dr^2+ g_{\mu\nu} (r,x^\mu )  dx^\mu dx^\nu
	+L^2(r,x^\mu)d\theta^2  \ . 
	\label{metric}
\end{eqnarray}
This assumption corresponds to the $Z_2$ symmetry in the 
 Randall-Sundrum braneworld model. 
Here we have introduced polar coordinates $(r,\theta)$ for 
the two extra spatial dimensions, where 
$0\leq r < \infty$ and $0 \leq\theta < 2\pi$. 
As we locate a four-dimensional brane at $r=0$, which is 
a string like defect, we must take the boundary condition
\begin{eqnarray}
 \lim_{r\rightarrow 0} L (r , x^\mu ) = 0 \ ,\qquad
 \lim_{r\rightarrow 0} L' (r , x^\mu ) = {\rm const.} 
 \label{bc:L}\ ,
\end{eqnarray}
where the prime denotes derivatives with respect to $r$.
The first condition realizes the 4-dimensional
brane at $r=0$. And the second condition
allows the existence of the conical singularity. 
The structure of conical singularity is a 2-dimensional delta 
function $\delta(r)/L$.
As we will see later,
the deficit angle is determined by the tension of the brane.   

 Possible components of the energy-momentum 
tensor $T_{ab}$, which could be balanced with the singular part of 
 Einstein tensor, are given by
\begin{eqnarray}
T_{ab}(\rm singular)=
	\left(\begin{array}{ccc}
	0&0&0\\
	0&T_{\mu\nu}({\rm singular})&0\\
	0&0&T_{\theta\theta}({\rm singular})\\
	\end{array}\right)  \ .
	\label{EM0}
\end{eqnarray}
Notice that $(r,r)$ component does not appear in the energy-momentum 
 tensor (\ref{EM0}), 
because this component must be balanced with the $(r,r)$ part of
 Einstein tensor 
 which consists of only the first derivatives with respect to $r$
 and hence there is no chance
 to have a singularity. On the other hand, $(\theta , \theta )$ component
 may have a singularity. 
 To determine the structure of the singularity in the  energy-momentum tensor, 
 we need to examine the structure of the singularity in the Einstein tensor.
 The singular part of the Einstein tensor gives the matching conditions.  
Using the metric (\ref{metric}), 
the non-linear matching condition for $(\theta, \theta)$
component is written as
\begin{eqnarray}
&&\hspace{-1cm}
	-K'+\alpha\bigg[ 
	\frac{4}{3}K^\alpha{}_\beta K^\beta{}_\gamma K^\gamma{}_\alpha
	-2KK^\beta{}_\alpha K^\alpha{}_\beta  
	+\frac{2}{3} K^3\bigg. \nonumber\\
&&\bigg.\qquad
	+4 K^\beta{}_\alpha R^\alpha{}_\beta -2 KR \bigg]' 
	=\kappa^2T^\theta{}_\theta~(\rm singular)\ ,
	\label{MC:tt}
\end{eqnarray}
where $K$ is the trace of is the extrinsic curvature, 
$K_{\mu\nu}=-1/2g_{\mu\nu,r}$. The $(\mu,\nu)$ components of
matching conditions read
\begin{eqnarray}
&&\hspace{-5mm}
	\delta^\mu_\nu \frac{L''}{L} 
	-4\alpha\bigg[ 
	\frac{L''}{L}G^\mu{}_\nu
	+\Big(\frac{L'}{L}W^\mu{}_\nu\Big)'
	\bigg]
	+\bigg[~
	K^\mu{}_\nu - \delta^\mu_\nu K~\bigg]'\nonumber\\
&&
	-4\alpha\bigg[~
	K^\mu{}_\nu \frac{\square L}{L} 
	+K\frac{L^{|\mu}{}_{|\nu}}{L}
	-K^\mu{}_\alpha\frac{L^{|\alpha}{}_{|\nu}}{L}  
	\bigg. \nonumber\\
&&\bigg.\qquad
	-K^\alpha{}_\nu \frac{L^{|\mu}{}_{|\alpha}}{L}
	-\delta^\mu_\nu\Big(~
	K\frac{\square L}{L}
	-K^\alpha{}_\beta\frac{L^{|\beta}{}_{|\alpha}}{L}~
	\Big)\bigg]' \nonumber\\
&&
	+2\alpha\bigg[~
	K^\mu{}_\nu K^\alpha{}_\beta K^\beta{}_\alpha
	-2K^\mu{}_\alpha K^\alpha{}_\beta K^\beta{}_\nu
	+2KK^\mu{}_\alpha K^\alpha{}_\nu
	\bigg.\nonumber\\
&&\left.\qquad
	-K^2 K^\mu{}_\nu
	-2K^\alpha{}_\beta R^{\mu\beta}{}_{\nu\alpha}
	+2KR^\mu{}_\nu
	\nonumber\right.\\
&&\left.\qquad
	-2K^\mu{}_\alpha R^\alpha{}_\nu
	-2K^\alpha{}_\nu R^\mu{}_\alpha
	+K^\mu{}_\nu R 
	\right.\nonumber\\
&& \left.\qquad 
	+\delta^\mu_\nu\Big(~ 
	\frac{2}{3}K^\alpha{}_\beta K^\beta{}_\gamma K^\gamma{}_\alpha
	-K K^\beta{}_\alpha K^\alpha{}_\beta +\frac{1}{3} K^3
	\Big.\right. \nonumber\\
&&\Big.\bigg.\qquad
	+2K^\beta{}_\alpha R^\alpha{}_\beta -KR~ 
	\Big) \bigg]^\prime 
	=\kappa^2T^\mu{}_\nu~(\rm singular)\ ,
	\label{MC:main}
\end{eqnarray}
where $|$ denotes the 4-dimensional covariant derivative and
 $W^{\mu}{}_\nu$ is defined as 
\begin{eqnarray}
W^\mu{}_\nu=K^\mu{}_\alpha K^\alpha{}_\nu
	-KK^\mu{}_\nu 
	+\frac{1}{2} \delta^\mu_\nu \left( K^2 
	-K^\alpha{}_\beta K^\beta{}_\alpha \right) \ .
	\label{MC:4d}
\end{eqnarray}
 In general, $K_{\mu\nu}$ and $L'/L$ could have discontinuities at $r=0$,
 namely, ${\displaystyle\lim_{\epsilon \rightarrow 0}} K_{\mu\nu} 
 (r=\epsilon , x^\mu)
 \neq K_{\mu\nu} (r=0 , x^\mu )$ and the same for $L'/L$. 
 For this case, two kinds of singular structures exist in 
Eqs.~(\ref{MC:tt}) and (\ref{MC:main}) under the boundary condition
(\ref{bc:L}).
One is a 2-dimensional delta function singularity 
$\delta(r)/2\pi L$ like $K'_{\mu\nu}/L$, $L''/L$, etc.  
The other is a 1-dimensional singularity $\delta(r)$ such as 
$K'_{\mu\nu}$, which is less singular. 
These kinds of singularities require the same kinds of singular
contribution in the energy-momentum tensor. 
The origin of two kinds of singularities in the energy-momentum tensor
 can be understood by the
 following schematic Taylor expansion:
\begin{eqnarray}
&&\hspace{-5mm}
{1\over \epsilon^2 }T_{ab} (r=\epsilon ,x^\mu ) \nonumber \\ 
&&\quad  = {1\over \epsilon^2 } \overset{(0)}{T}{}_{ab} (r=0, x^\mu ) 
   + {1\over \epsilon }\overset{(1)}{T}{}_{ab} (r=0, x^\mu ) \ , 
\end{eqnarray}
where  $T_{ab}$ is the energy-momentum tensor of smooth matter, and 
 $\overset{(0)}{T}{}_{ab}$
 and $\overset{(1)}{T}{}_{ab}$ denote the  Taylor coefficients.
 Here, $\epsilon$ represents the thickness of the matter distribution.
In the conventional thin limit, only the first term is usually taken.
 We propose to incorporate the second term 
  to take into account the thickness of braneworld. 
  Thus the possible energy-momentum tensor takes the form
\begin{eqnarray}
T_{ab} ({\rm singular})=
	\left(\begin{array}{ccc}
	0&0&0\\
	0&T_{\mu\nu}\frac{\delta(r)}{2\pi L}+S_{\mu\nu}
	\delta(r)&0\\
	0&0&S_{\theta\theta}\delta(r)\\
	\end{array}\right)  \ ,
	\label{EM}
\end{eqnarray}
where $T_{\mu\nu}$ represents the conventional brane matter, 
while $S_{\mu\nu}$ and $S_{\theta\theta}$ describe
 the extra matter which mimics the thickness of the braneworld.
 Note that $T_{\theta\theta}$ can not be allowed due to the structure
 of Eq.~(\ref{MC:tt}). 
 It should be stressed that this ansatz 
 completely makes sense in the axially symmetric spacetime. 
 In this way, the thickness can be treated  within the context of a 
 distributional source, which we dub quasi-thickness.
 In a sense, we are considering the ``internal structure" of 
 the braneworld.

\section{Vacuum Braneworld}

 The linear analysis is a first step to understand the effective
 theory on the braneworld. As a background spacetime, 
 let us first consider the vacuum braneworld. Later, we will
 consider the fluctuations around this vacuum solution.
 Then, we can proceed to the discussion of the nonlinear dynamics of
 the codimension 2 braneworld.

  Because of the symmetry  of the vacuum, the metric can be expressed as
\begin{eqnarray}
ds^2=dr^2+a^2(r)\eta_{\mu\nu}dx^\mu dx^\nu
	+b^2(r)d\theta^2 \ ,
	\label{bg}
\end{eqnarray}
where $a(r)$ and $b(r)$ depend only on $r$.
The energy-momentum tensor (\ref{EM}) of the vacuum brane can 
be characterized by the tension $\sigma$:
\begin{eqnarray}
T^{a}{}_{b} ({\rm singular})=
	\left(\begin{array}{ccc}
	0&0&0\\
	0&-\sigma\delta^\mu_\nu\frac{\delta(r)}{2\pi b}&0\\
	0&0&0\\
	\end{array}\right)  \ .
\end{eqnarray}
Here, we have introduced only the 2-dimensional distribution.
 This seems a reasonable assumption because the vacuum can not have
 an internal structure.

Given the  metric (\ref{bg}), the Einstein Gauss-Bonnet equations~(\ref{EGB}), 
off the brane, in the bulk are written as
\begin{eqnarray}
&&\hspace{-5mm}
3\frac{a''}{a}+3\left(\frac{a'}{a}\right)^2
	+3\frac{a'b'}{ab}+\frac{b''}{b} \nonumber\\
&&
	-12\alpha\left[\frac{a''a'^2}{a^3}
      	+\frac{a'^3b'}{a^3b}
      	+2\frac{a''a'b'}{a^2b}
      	+\frac{a'^2 b''}{a^2b}
      	\right]=0
      	\label{munu}\ ,\\
&&\hspace{-5mm}
6\left(\frac{a'}{a}\right)^2+4\frac{a'b'}{ab}
	-12\alpha\left[\left(\frac{a'}{a}\right)^4
	+4\frac{a'^3b'}{a^3b}\right]=0\ ,
	\\
&&\hspace{-5mm}
4\frac{a''}{a}+6\left(\frac{a'}{a}\right)^2
	-12\alpha\left[\left(\frac{a'}{a}\right)^4
	+4\frac{a''a'^2}{a^3}
	\right]=0 \ .
\end{eqnarray}
The solution  is obtained as 
\begin{eqnarray}
ds^2=dr^2+ \eta_{\mu\nu}dx^\mu dx^\nu
	+c^2r^2d\theta^2\ ,
\end{eqnarray}
where $c$ is a constant of integration. For $c\ne 1$, we have
a conical singularity at $r=0$.
This  can be seen in the following way:
 First, we specify the boundary condition at $r=0$
 by following the standard procedure, $b' (r=0 ) =1$.
 Then, the 2-dimensional curvature is calculated as
\begin{eqnarray}
\frac{b''}{b} \Big|_{r=0}
         &=& \lim_{\epsilon \rightarrow 0}\frac{1}{b} 
   \left( \frac{b'(r=\epsilon )-b'(r=0)}{\epsilon}\right) \nonumber\\
         &=& (c-1) \frac{\delta (r)}{b}   \ . 
\end{eqnarray}

 Now, the matching condition (\ref{MC:main}) 
determines the deficit angle in terms of the brane tension as
\begin{eqnarray}
1-c =\kappa^2 {\sigma \over 2\pi} \ .
\label{mc}
\end{eqnarray}
Note that $c < 1$, because the tension $\sigma$ is positive.
This result gives the insight that the deficit might depend on $x^\mu$
 when the inhomogeneous matter exists on the brane.

\section{Linear Perturbation Analysis}

Having considered a background solution and found that the deficit 
angle seems to depend on the brane matter, we now consider a 
perturbation around this solution to examine the behavior of the 
deficit angle fluctuations. 

 It is possible to classify perturbations using properties
 under the 4-dimensional coordinate transformation.
 Let us discuss the scalar, vector, and tensor perturbations, separately.

\subsection{Scalar Perturbation}

 Although we  take the Gaussian normal coordinate 
  system for discussing the effective theory on the brane, 
 the following gauge is convenient for the 
calculation in the case of the scalar perturbations: 
\begin{eqnarray}
ds^2&=&(1+\delta\varphi+2\Psi)dr^2
	+(1+\delta\varphi+2\Phi)\eta_{\mu\nu}dx^\mu dx^\nu
	\nonumber\\
	&&\qquad
	+r^2(1-3\delta\varphi)d\theta^2 \ ,
	\label{sp}
\end{eqnarray}
where we absorbed the background value of the deficit angle
 into the definition of $\theta$, i.e. $0 \leq\theta < 2\pi c$.
In linearized Einstein Gauss-Bonnet gravity, we do not require the
GB term in the flat bulk background because it consists of
a quadratic form of curvature. Then, the 6-dimensional Einstein equation
becomes

\begin{eqnarray}
&&\hspace{-5mm}
\frac{1}{2}\square\delta\varphi+\frac{1}{2}\delta\varphi''
	-\frac{7}{2r}\delta\varphi'+\square\Psi
	-\frac{1}{r}\Psi'+4\Phi''=0\label{sp:rr}\ ,\\
&&\hspace{-5mm}
\Big[-3\Phi'+\frac{2}{r}\delta\varphi+\frac{1}{r}\Psi\Big]_{,\mu}=0
	\label{sp:rmu}\ ,\\
&&\hspace{-5mm}
\frac{3}{2}\square\delta\varphi+\frac{3}{2}\delta\varphi''
	+\frac{3}{2r}\delta\varphi'+\frac{1}{r}\Psi'
	-\frac{4}{r}\Phi'=0\label{sp:tt}\ ,\\
&&\hspace{-5mm}
\Big[\Psi+2\Phi\Big]_{,\mu\nu}=0\label{sp:trl}\ ,\\
&&\hspace{-5mm}
\frac{1}{2}\square\delta\varphi+\frac{1}{2}\delta\varphi''
	+\frac{1}{2r}\delta\varphi'+\square\Phi+\Phi''
	+\frac{1}{r}\Phi'=0\label{sp:tr}\ .
\end{eqnarray}
Integrating Eqs.~(\ref{sp:rmu}) and (\ref{sp:trl}), we can 
 write $\Psi$ and $\delta\varphi$  in terms of $\Phi$,
\begin{eqnarray}
&&\Psi~=-2\Phi\label{psi}\ ,\\
&&\delta\varphi=\frac{3}{2}r\Phi'+\Phi+\frac{r}{2}d(r)
\label{varphi}\ ,
\end{eqnarray}
where $d(r)$ is a constant of integration and depends only on $r$. 
Here, we imposed the regularity at $x^\mu\rightarrow\infty$. 
After eliminating $\square\delta\phi$ using Eqs.~(\ref{sp:tt}) and 
(\ref{sp:tr}), putting Eqs.~(\ref{psi}) and (\ref{varphi}) in the 
resulting equation, we find the equation of motion
for $\Phi$, 
\begin{eqnarray}
\Phi''+\frac{3}{r}\Phi'+\square\Phi=0\ .
\label{eom:phi}
\end{eqnarray}
 After applying the same procedure to Eq.~(\ref{sp:rr}) and 
(\ref{sp:tr}), we obtain
\begin{eqnarray}
\Phi''+\frac{3}{r}\Phi'+\square\Phi +{2\over 3 r} [d(r)r]' =0 \ .
\label{eom:phiprime}
\end{eqnarray}
Comparing the above Eq.~(\ref{eom:phiprime}) with Eq.~(\ref{eom:phi}), 
 we get 
\begin{eqnarray}
d(r)r={\rm const.} \equiv C_0     \ .
\end{eqnarray}
This constant is related to the mass of the system
 determined from the given energy-momentum tensor~\cite{Frolov}.

If we define Fourier transformation,
\begin{eqnarray}
\Phi(r,x^\mu)=\int d^4p e^{ip\cdot x}\Phi(r,p) \ ,
\end{eqnarray}
Then the solution of Eq.~(\ref{eom:phi}) is
\begin{eqnarray}
\Phi(r,p)=\frac{A(q)}{r}J_1(qr) \ ,
\end{eqnarray}
where $q^2=-p^\mu p_\mu$ and we have imposed the regularity at the origin.
 The amplitude $A(q)$ should be determined by the matching condition.  
The solution is given by 
\begin{eqnarray}
ds^2&=&\left(  1+\delta g_{rr} \right) dr^2
	+\left(\eta_{\mu\nu}+\delta g_{\mu\nu} \right)
	dx^\mu dx^\nu \nonumber\\
 && \quad+\left(r^2 + \delta g_{\theta \theta}\right)d\theta^2\ ,
\end{eqnarray}
where
\begin{eqnarray} 
	 \delta g_{rr}&=&\frac{C_0}{2}-\frac{3}{r}Z_1(r,x^\mu)
	-\frac{3}{2}Z_2(r,x^\mu) \ , \\
	 \delta g_{\mu\nu} &=& \left( \frac{C_0}{2}
	+\frac{3}{r}Z_1(r,x^\mu)
	-\frac{3}{2}Z_2(r,x^\mu) \right) \eta_{\mu\nu} \ ,\\
	 \delta g_{\theta \theta} &=& r^2\left(-\frac{3}{2}C_0
	-\frac{3}{r}Z_1(r,x^\mu)
	+\frac{9}{2}Z_2(r,x^\mu)\right) \ .
\end{eqnarray}
Here, we have defined
\begin{eqnarray}
Z_1(r,x^\mu)&=&\int d^4p~e^{ip\cdot x}A(q)J_1(qr)\ ,\\
Z_2(r,x^\mu)&=&\int d^4p~e^{ip\cdot x}A(q)qJ_2(qr)\ .
\end{eqnarray}

In order to impose the matching conditions, we must move to the 
Gaussian normal coordinate system
\begin{eqnarray}
  ds^2 = dr^2
	+\left(\eta_{\mu\nu}+\delta \bar{g}_{\mu\nu} \right)
	dx^\mu dx^\nu 
	+\left(1+ \delta \bar{g}_{\theta \theta}\right)d\theta^2 
	\label{GN:general}\ ,
\end{eqnarray}
where the axial symmetry is taken into account.
This can be achieved by the gauge transformations associated with
 the infinitesimal coordinate transformations 
 $x^\mu \rightarrow x^\mu - \xi^\mu $:
\begin{eqnarray}
\delta \bar{g}_{\mu\nu} &=& \delta g_{\mu\nu} 
           + \xi_{\mu ,\nu} + \xi_{\nu ,\mu}\ , \\
\delta \bar{g}_{\theta \theta} &=& \delta g_{\theta \theta}
           + 2 c^2 r \xi_r\ ,           
\end{eqnarray}
where
\begin{eqnarray}
    \xi_r &=& -{1\over 2} \int^r_0 dr \delta g_{rr} \ , \\
    \xi_\mu &=& {1\over 2} \int^r_0 dr \int^r_0 dr \delta g_{rr , \mu} 
                                        \ .
\end{eqnarray}
Note that the location of the brane is not necessarily at constant $r$.
 However, there exist residual gauge transformations
\begin{eqnarray}
    \xi_r &=&  \chi (x^\mu )\ , \\
    \xi_\mu &=&  \chi_{,\mu}\ r  \ .
\end{eqnarray}
Using this residual gauge, one can adjust the brane position to be 
 located at the constant $r$ in the new coordinate system. 
 Thus, the distortion function $\chi(x^\mu)$ takes into account
  the fact that the brane is not necessarily
 located at the constant $r$ in the coordinate system 
which we used to solve equations in the bulk. 
The final result is written by
\begin{eqnarray}
ds^2&=&  dr^2
	+\left(1+h_{\mu\nu}\right)
	\eta_{\mu\nu}dx^\mu dx^\nu
	+L^2d\theta^2\ ,
\end{eqnarray}
where
\begin{eqnarray}
h_{\mu\nu}&=&\frac{C_0}{2}+\frac{3}{r}Z_1(r,x^\mu)
	-\frac{3}{2}Z_2(r,x^\mu)\nonumber\\
&&	+\int^r_0dr \int^r_0dr\delta g_{rr ,\mu\nu}
	-2\chi_{,\mu\nu}~r\ ,   \label{scalar:metric}  \\
L^2&=&c^2r^2\left[1-\frac{3}{2}C_0
	-\frac{3}{r}Z_1(r,x^\mu)
	+\frac{9}{2}Z_2(r,x^\mu)\right]
	\nonumber\\
&&	- c^2 r \int^r_0dr\delta g_{rr}
	+2c^2r\chi\ , \label{scalar:L}
\end{eqnarray}
and we have reintroduced the background deficit, $c$, explicitly.
Now, the brane is located at $r=0$ in this coordinate system.

\subsection{Vector perturbations}

Now, we move on to vector perturbations which turn out to be irrelevant. 
As to the vector perturbations,  we choose the gauge 
\begin{eqnarray}
ds^2=dr^2+\left(\eta_{\mu\nu}+F_{\mu,\nu}+F_{\nu,\mu}\right)dx^\mu dx^\nu
	+c^2r^2d\theta^2 \ ,
\end{eqnarray}
where $F^\mu{}_{,\mu} = 0$.
Note that there exist residual gauge transformations
\begin{eqnarray}
  \bar{F}_\mu = F_\mu + \eta_\mu (x^\mu )\ ,\qquad\eta^\mu{}_{,\mu} =0 \ ,
  \label{gaugetrans}
\end{eqnarray} 
where the parameter $\eta_\mu$  depends only on $x^\mu$.
  
The equation of motion for $F_\mu$ becomes
\begin{eqnarray}
\left(F_{\mu,\nu}+F_{\nu,\mu}\right)''
	+\frac{1}{r}\left(F_{\mu,\nu}+F_{\nu,\mu}\right)'
	=0 \ .
\end{eqnarray}
The regular solution of this equation is
\begin{eqnarray}
F_\mu=f_{\mu}(x^\mu) \ .
\end{eqnarray}
This solution can be eliminated using the residual 
gauge transformations (\ref{gaugetrans}). 
Hence,  there exist no vector perturbations.

\subsection{Tensor perturbations}

Finally, we  consider tensor perturbations. 
They are characterized by a metric
\begin{eqnarray}
ds^2=dr^2+\left(\eta_{\mu\nu}+h^{\rm TT}_{\mu\nu}\right)dx^\mu dx^\nu
	+c^2r^2d\theta^2 \ ,
\end{eqnarray}
where $h^{\rm TT}_{\mu\nu}$ satisfies transverse and traceless conditions.
Tensor perturbations remain invariant
under a coordinate transformation.
The equation of motion for $h^{\rm TT}_{\mu\nu}$ gives
\begin{eqnarray}
h^{\prime\prime{\rm TT}}_{\mu\nu}+\frac{1}{r}h^{\prime\rm TT}_{\mu\nu}
	+\square h^{\rm TT}_{\mu\nu}=0\ .
\end{eqnarray}
The solution becomes
\begin{eqnarray}
h^{\rm TT}_{\mu\nu}(r,x^\mu)=
\int d^4p~e^{ip\cdot x}\varepsilon_{\mu\nu}(q)J_0(qr) \ ,
  \label{tensormode}
\end{eqnarray}
where $\varepsilon_{\mu\nu}$ should be determined 
 by the matching condition.

\section{Effective Theory}

%
\subsection{Linear Regime}

 Since we have solved the bulk geometry, we next discuss the 
 matching conditions. Using the energy-momentum tensor (\ref{EM}),
matching conditions Eqs.~(\ref{MC:tt}) and (\ref{MC:main}) reduce to
\begin{eqnarray}
&&-\delta K^\prime=\kappa^2S^{\theta}{}_{\theta}\delta\big(r\big)
\label{MC1:tt}\ ,\\
&&\delta K^{\prime\mu}{}_{\nu}-\delta^\mu_\nu \delta K'
	=\kappa^2S^{\mu}{}_{\nu}\delta\big(r\big)
	\label{MC1:4d}\ ,\\
&&\delta^\mu_\nu\delta L''
	-4\alpha L''\delta G^{\mu}{}_{\nu}
	=\frac{\kappa^2}{2\pi}T^{\mu}{}_{\nu}\delta(r)
	\label{MC1:main}\ ,
\end{eqnarray}
where we dropped all the remaining terms in Eqs.~(\ref{MC:tt}) 
and (\ref{MC:main}), because they do not contribute at first order 
in perturbation. 

 Using Eq.~(\ref{scalar:L}),
 the deficit angle can be obtained from
\begin{eqnarray}
 \lim_{\epsilon \rightarrow 0}
  L' (r =\epsilon ) &=& \lim_{\epsilon \rightarrow 0}
  \left[ b'(r=\epsilon)+\delta L'(r=\epsilon) \right] \nonumber\\
 &=& c(1-C_0)\ .
\end{eqnarray}
Note that the deficit angle is  perturbed on the brane 
at the linear level, but it is constant. 
 From Eqs.~(\ref{scalar:metric}) and (\ref{tensormode}),
 the extrinsic curvature in this limit gives 
\begin{eqnarray}
 \lim_{\epsilon \rightarrow 0} 
  K_{\mu\nu}(r=\epsilon,x^\mu)&=&
 \lim_{\epsilon \rightarrow 0} 
 \delta K_{\mu\nu}(r=\epsilon,x^\mu)  \nonumber\\
 &=& -\frac{1}{2}\lim_{\epsilon \rightarrow 0} 
 h'_{\mu\nu} (r=\epsilon) =\chi_{,\mu\nu} \ .
        \label{Ke}
\end{eqnarray}
Notice that the background value of $K_{\mu\nu}$ is zero 
 and the tensor perturbations do not contribute to
 the extrinsic curvature (\ref{Ke}).

 From this result,
we see that the extrinsic curvature is nonzero near the brane, 
if the distortion field $\chi (x^\mu)$ exists.
 In the coordinate system where we have solved the equations
 of motion in the bulk, the location of the brane is specified by $\chi (x^\mu)$
 and the resulting shape of the braneworld becomes the deformed cylinder
 in which the distortion is represented by $\chi (x^\mu)$ (see Fig.1). 
 At first sight, this deformed cylinder looks like a 5-dimensional
 hypersurface. However,
 at the location of the brane ($r=\epsilon$ in the Gaussian normal gauge),  
 the circumference radius of the cylinder vanishes because of 
 ${\displaystyle\lim_{\epsilon \rightarrow 0}} L(r=\epsilon ,x^\mu) =0 $ there.
 Hence, the deformed cylinder is the 4-dimensional braneworld. 
\begin{figure}[h]
\centerline{\includegraphics[height=6cm, width=5.5cm]{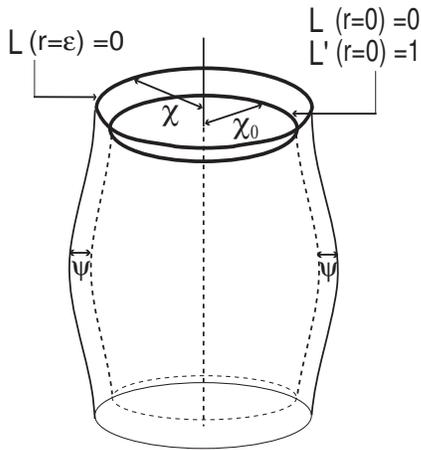}}
\caption{Schematic picture for the quasi-thick braneworld. 
The deformed cylinder has zero circumference radius. 
Hence, it is a 4-dimensional
 spacetime. $\psi (x^\mu) = \chi (x^\mu ) -\chi_0 (x^\mu )$ 
 represents the thickness of the brane. }
\end{figure}

 Now, we must specify $K_{\mu\nu}(r=0,x^\mu )$ which is not known a priori.
 Without losing the generality, we can write
\begin{eqnarray}
  K_{\mu\nu}(r=0,x^\mu ) = \chi_{0,\mu\nu} (x^\mu ) + \rho_{\mu\nu} (x^\mu )
  \ , \quad \rho^\mu{}_\mu =0 \ , 
\end{eqnarray}
where we decomposed the extrinsic curvature at $r=0$ into the traceless part
 $\rho_{\mu\nu} (x^\mu)$ and the trace part $ \square \chi_0 (x^\mu )$. 
 As the field $\chi (x^\mu )$ represents the location of the brane 
 $r=\epsilon $, it is natural to regard $\chi_0 (x^\mu )$ 
 as the location of $r=0$. 
 Hence,  $\psi (x^\mu ) \equiv \chi (x^\mu ) -\chi_0 (x^\mu )$ 
 can be interpreted as  the thickness of the braneworld.

 With these considerations, we find Eqs.~(\ref{MC1:tt}), (\ref{MC1:4d}) 
 and (\ref{MC1:main}) become
\begin{eqnarray}
&&\hspace{-5mm}
\square(\chi-\chi_0)=-\kappa^2S^\theta{}_\theta
\label{mc:tt}\ ,\\
&&\hspace{-5mm}
\chi^{,\mu}{}_{,\nu}-\delta^\mu_\nu\square\chi  
	-\left[\chi_0{}^{,\mu}{}_{,\nu} 
	-\delta^\mu_\nu\square\chi_0\right] 
	-\rho^\mu{}_\nu
	=\kappa^2S^\mu{}_\nu
	\label{mc:4d}\ ,\\
&&\hspace{-5mm}
\delta^\mu_\nu\left[
	\delta L'-\delta L_0'\right]-4\alpha
	\left[
	b'-b_0'\right]\delta G^\mu{}_{\nu}
	=\frac{\kappa^2}{2\pi}T^\mu{}_\nu\ .
	\label{MC1:Main}
\end{eqnarray}
Adopting the standard boundary condition 
$L' (r=0) =b'(r=0)+\delta L'(r=0)=1$  even in the perturbed spacetime, 
Eq.~(\ref{MC1:Main}) leads to
\begin{eqnarray}
G^\mu{}_\nu =\frac{\kappa^2}{8\pi\alpha(1-c)}T^{\mu}{}_{\nu}
	+\delta^\mu_\nu\frac{cC_0}{4\alpha(1-c)}  \ ,
	\label{main}
\end{eqnarray}
 where the second term of the right hand side of (\ref{main})
 is the cosmological constant.
 The Einstein tensor includes the contribution from the tensor perturbations. 
 Eq.~(\ref{main}) proves that Einstein gravity is recovered for 
 the codimension 2 braneworld in the case of the linearized gravity.
 This fact  is consistent with the result in \cite{Bostock}.
 However, this is not the end of our story. 
 We must also solve the conditions (\ref{mc:tt}) and (\ref{mc:4d}).

The conservation law $T^a{}_{e;a}=0$ for the $e=\mu$ at the 
linearized level gives
the 4-dimensional conservation law,
\begin{eqnarray}
T^{\mu\nu}{}_{|\nu} = 0 \ ,\qquad
S^{\mu\nu}{}_{|\nu} = 0 \ ,
\end{eqnarray}
where $|$ denotes the 4-dimensional covariant derivative.
On the other hand, the conservation law for the $e=r$ implies,
\begin{eqnarray}
\left[\delta K\frac{\sigma}{2\pi} + b' S^\theta{}_\theta \right]
\delta (r)=0\ .
\label{cl:r}
\end{eqnarray}
Note that Eq.~(\ref{cl:r}) appears to contain an ambiguity, since
$K_{\mu\nu}$ must be evaluated at $r=0$ where it is discontinuous.
Thus we define the following quantities~\cite{Blau}
\begin{eqnarray}
\delta\bar{K}&=&\lim_{\epsilon\rightarrow 0}\frac{1}{2}
	\left[\delta K(r=\epsilon)
	+\delta K(r=0)\right]\ ,\\
\bar{b}^\prime&=&\lim_{\epsilon\rightarrow 0}\frac{1}{2}
	\left[ b^\prime (r=\epsilon)
	             + b^\prime(r=0)\right]\ .
\end{eqnarray}
Using these quantities, one can deduce 
\begin{eqnarray}
\delta\bar{K}\frac{\sigma}{2\pi}+\bar{b}^\prime S^\theta{}_\theta
 	=0 \ .
 	\label{cl:r2}
\end{eqnarray}
 The conservation law (\ref{cl:r2}) gives the constraint
on the distributional sources as
\begin{eqnarray}
\square(\chi+\chi_0)=-\frac{1+c}{1-c}\kappa^2S^\theta{}_\theta
\label{r:cl}\ ,
\end{eqnarray}
where we used Eq.~(\ref{mc}). Combining  Eq.~(\ref{mc:tt}) with 
(\ref{r:cl}), we find
\begin{eqnarray}
\square\chi=-\frac{\kappa^2}{1-c}S^\theta{}_\theta,\qquad
\square\chi_0=-\frac{\kappa^2c}{1-c}S^\theta{}_\theta \ .
 \label{chi:eq}
\end{eqnarray}
The above Eqs.~(\ref{chi:eq})  tell us that $\chi_0 = c \chi$.
 Since $c<1$,  $\psi=\chi-\chi_0= (1-c)\chi$ is positive.
 Hence, it is legitimate to interpret $\psi$  as the effective thickness.  
 In the meantime, 
 comparing  Eq.~(\ref{mc:tt}) with the trace of Eq.~(\ref{mc:4d}), 
 we  find the relation between the components of 
energy-momentum tensor as
\begin{eqnarray}
S^\theta{}_\theta=\frac{1}{3}S^\mu{}_\mu \ .
\end{eqnarray}
 Hence, once we give the 4-dimensional matter $S_{\mu\nu}$, 
 the extra component $S_{\theta\theta}$ can be determined. 
 It should be noted that the effective thickness $\psi$ satisfies
\begin{eqnarray}
\square\psi=-\frac{\kappa^2}{3}S^\mu{}_\mu\ .
\end{eqnarray}
This is reminiscent of the radion in the codimension 1 
braneworld~\cite{Garriga:1999yh}.
 The remaining quantity $\rho_{\mu\nu}$ can be also obtained 
 from Eq.~(\ref{mc:4d}). 

%
\subsection{Nonlinear Regime}

 Our main point is the formulation of the framework of the
 quasi-thick braneworld. The importance of this point can be
 manifest in the next order calculations. 
Up to the second order, we expect 
\begin{eqnarray}
G^\mu{}_\nu&=&\frac{\kappa^2}{8\pi\alpha (1-c+cC_0)}T^\mu{}_\nu 
	+\frac{cC_0-\overset{(2)}{L}{}^{\prime}}{4\alpha (1-c+cC_0)}
	\delta^\mu_\nu
	\nonumber\\
&&
	+ c\Big[\chi^{,\mu}{}_{,\alpha} \chi^{,\alpha}{}_{,\nu}
	- \square \chi \chi^{, \mu}{}_{, \nu}\Big.\nonumber\\
&&\Big.\qquad
	+\frac{1}{2} \delta^\mu_\nu \left\{ (\square\chi)^2 
	-\chi^{,\alpha}{}_{,\beta} \chi^{,\beta}{}_{,\alpha}\right\}
	\Big] \nonumber \\
&&	-\frac{1}{1-c} \left[ \rho^\mu{}_\alpha \rho^\alpha{}_\nu
	- \delta^\mu_\nu \rho^\alpha{}_\beta \rho^\beta{}_\alpha
	      \right]   \label{second-order} \ ,
\end{eqnarray}
where the corrections in the last three lines arises due to  
$\overset{(2)}{W}{}_{\mu\nu}$
and carry the information of thickness of the brane. Another correction
come from $\overset{(2)}{L}{}^{\prime}$ carries the effects of
bulk, that is, the deformed deficit angle. This needs the next
order analysis. Thus, it turns out that some corrections due 
to the quasi-thickness can be expected at the second order. 
 
 Here, we should point out a possibility that the above corrections
to Einstein equations could mimic the observed features of dark
energy/dark matter. To give some flavor, let us extrapolate the effective theory
 (\ref{second-order}) to the nonlinear regime. 
In the case of cosmology, one can put 
\begin{eqnarray}
S^{\mu}{}_{\nu}=
	\left(\begin{array}{cc}
	-\rho_{\rm D}&0\\
	0&p_{\rm D}\delta^i_j\\
	\end{array}\right)  \ .
\end{eqnarray}
Then, one finds when the condition
\begin{eqnarray}
\frac{5\kappa^4}{3(1-c)}\rho_{\rm D}^2
	< \frac{\overset{(2)}{L}{}^{\prime} -cC_0}{4\alpha (1-c+cC_0)}
\end{eqnarray}
is satisfied, these kinds of corrections behave as dark energy.
While, when the condition
\begin{eqnarray}
\frac{\overset{(2)}{L}{}^{\prime}-cC_0}{4\alpha (1-c+cC_0)}
	=\frac{11\kappa^4}{9(1-c)}\rho_{\rm D}^2
\end{eqnarray}
is satisfied, the effective energy-momentum tensor due to corrections
 can be regarded as dark matter on the brane. 
 To verify these expectations, we need to solve the bulk geometry.

 In reality, we need to consider the full nonlinear theory to discuss
 the cosmology. The nonlinear effective equation should read
\begin{eqnarray}
&&\hspace{-5mm}
	\lim_{\epsilon\rightarrow 0}
	\bigg[ \delta^\mu_\nu L'
	-4\alpha\Big\{ 
	L' G^\mu{}_\nu
	+ L' W^\mu{}_\nu
	\Big\}   \bigg. \nonumber\\
&& \bigg.
	-4\alpha\Big\{
	K^\mu{}_\nu \square L 
	+K   L^{|\mu}{}_{|\nu}
	-K^\mu{}_\alpha  L^{|\alpha}{}_{|\nu}  	
	-K^\alpha{}_\nu  L^{|\mu}{}_{|\alpha}
		\Big. \bigg. \nonumber\\
&& \bigg.\Big.\qquad
	-\delta^\mu_\nu\Big(~
	K \square L
	-K^\alpha{}_\beta  L^{|\beta}{}_{|\alpha}~
	\Big)\Big\} \bigg]\bigg|^{\epsilon}_0 
	= {\kappa^2 \over 2\pi} T_{\mu\nu}
	\label{nonlinear} \ ,
\end{eqnarray}
where the left hand side Eq.~(\ref{nonlinear}) 
represents the discontinuity
 at $r=0$. To fully solve the system consistently, we must also
 take into account the matching conditions come from
 the less singular part which determines the boundary conditions
 completely.

 It is also intriguing to see how the gravitational waves on the brane
 are affected by the bulk geometry in the nonlinear situations
 such as the radiation from the particle falling into the black hole.
 For this purpose, we must solve the above non-linear problem.


\section{Conclusion}

We have studied the codimension 2 Einstein Gauss-Bonnet braneworld 
in the axially symmetric 6-dimensional spacetime. 
 We  have carefully examined the structure of possible singularities
 in the equations of motion which characterize the braneworld through 
  matching conditions. It turned out that the
  thickness of the brane can be taken into account within the context of
  the distributional source, which we dubbed quasi-thickness.
 In the case of the linearized gravity,  we have solved all of the equations 
 of motion in the bulk and shown the conventional Einstein gravity 
 can be recovered on the brane.
 In this process, all of the necessary boundary conditions are clarified.
 In the nonlinear regime, we found corrections due to the thickness 
 and the bulk geometry. We stressed that the interplay between the bulk 
 and the brane has been completely determined in the context of
 the  quasi-thick braneworld. 
 
 In our formulation of the codimension 2 braneworld, $T_{\mu\nu}$
 can be identified with the ordinary matter. While $S_{\mu\nu}$
 is a kind of dark energy/dark matter. The possibility that the thickness plays
 a role of the dark sector in the universe is attractive.  
 It would be also interesting to construct
 a viable particle physics model in this context. 

 There are many remaining issues to be solved.
 Once $T_{\mu\nu}$ and $S_{\mu\nu}$ are given, we can analyze the nonlinear
 gravity to reveal the nature of the corrections. 
 This would be very complicated but the analysis is straightforward.
 Application to cosmology seems to be trivial in the linear regime.  This is 
 in contrast to the case of codimension 1 braneworld where  the analysis of 
 the cosmological perturbations are very complicated~\cite{soda}.  
 It is interesting to consider the black holes in the context of the 
 codimension 2 braneworld. The role of the Gregory-Laflamme instability
 should be clarified and the property of the gravitational waves from the
 black hole must be analyzed~\cite{GL}. 
 It is also intriguing to extend our formalism to the higher codimension
 braneworld~\cite{higher}.

\begin{acknowledgements}
This work was supported in part by  Grant-in-Aid for  Scientific
Research Fund of the Ministry of Education, Science and Culture of Japan 
 No. 155476 (SK) and  No.14540258 (JS) and also
  by a Grant-in-Aid for the 21st Century COE ``Center for
  Diversity and Universality in Physics".  
\end{acknowledgements}

%


\begin{thebibliography}{99}

\bibitem{Rubakov}
V.~A.~Rubakov and M.~E.~Shaposhnikov,
Phys.\ Lett.\ B {\bf 125}, 139 (1983);
V.~A.~Rubakov and M.~E.~Shaposhnikov,
Phys.\ Lett.\ B {\bf 125}, 136 (1983);
K.~Akama,
Lect.\ Notes Phys.\  {\bf 176}, 267 (1982)
[arXiv:hep-th/0001113].

\bibitem{Polchinski}
J.~Polchinski, String Theory I and II (Cambridge Univ. Press, Cambridge, 1998).

\bibitem{RS1}
L.~Randall and R.~Sundrum,
Phys.\ Rev.\ Lett.\  {\bf 83}, 3370 (1999)
[arXiv:hep-ph/9905221];
L.~Randall and R.~Sundrum,
Phys.\ Rev.\ Lett.\  {\bf 83}, 4690 (1999)
[arXiv:hep-th/9906064];
 see also the review paper and refereces therein:
 P.~Brax, C.~van de Bruck and A.~C.~Davis,
arXiv:hep-th/0404011.

\bibitem{Geroch}
R.~Geroch and J.~H.~Traschen,
Phys.\ Rev.\ D {\bf 36}, 1017 (1987).

\bibitem{codimension2}
T.~Gherghetta and M.~E.~Shaposhnikov,
Phys.\ Rev.\ Lett.\  {\bf 85}, 240 (2000)
[arXiv:hep-th/0004014];
I.~Navarro,
JCAP {\bf 0309}, 004 (2003)
[arXiv:hep-th/0302129];
I.~Navarro,
Class.\ Quant.\ Grav.\  {\bf 20}, 3603 (2003)
[arXiv:hep-th/0305014];
J.~M.~Cline, J.~Descheneau, M.~Giovannini and J.~Vinet,
JHEP {\bf 0306}, 048 (2003)
[arXiv:hep-th/0304147];
Y.~Aghababaie {\it et al.},
JHEP {\bf 0309}, 037 (2003)
[arXiv:hep-th/0308064];
H.~M.~Lee and G.~Tasinato,
arXiv:hep-th/0401221;
N.~Kaloper,
arXiv:hep-th/0403208;
I.~P.~Neupane,
Class.\ Quant.\ Grav.\  {\bf 19}, 5507 (2002)
[arXiv:hep-th/0106100];
Y.~M.~Cho and I.~P.~Neupane,
Int.\ J.\ Mod.\ Phys.\ A {\bf 18}, 2703 (2003)
[arXiv:hep-th/0112227];
O.~Corradini, A.~Iglesias, Z.~Kakushadze and P.~Langfelder,
Phys.\ Lett.\ B {\bf 521}, 96 (2001)
[arXiv:hep-th/0108055];
O.~Corradini, A.~Iglesias, Z.~Kakushadze and P.~Langfelder,
Mod.\ Phys.\ Lett.\ A {\bf 17}, 795 (2002)
[arXiv:hep-th/0201201];
O.~Corradini, A.~Iglesias and Z.~Kakushadze,
Int.\ J.\ Mod.\ Phys.\ A {\bf 18}, 3221 (2003)
[arXiv:hep-th/0212101];
O.~Corradini, A.~Iglesias and Z.~Kakushadze,
Mod.\ Phys.\ Lett.\ A {\bf 18}, 1343 (2003)
[arXiv:hep-th/0305164];
J.~E.~Kim, B.~Kyae and H.~M.~Lee,
Phys.\ Rev.\ D {\bf 64}, 065011 (2001)
[arXiv:hep-th/0104150].

\bibitem{Deruelle}
D.~Lovelock,
J.\ Math.\ Phys.\  {\bf 12}, 498 (1971);
N.~Deruelle and J.~Madore,
arXiv:gr-qc/0305004.

\bibitem{Bostock}
P.~Bostock, R.~Gregory, I.~Navarro and J.~Santiago,
arXiv:hep-th/0311074.

\bibitem{Frolov}
V.~P.~Frolov, W.~Israel and W.~G.~Unruh,
Phys.\ Rev.\ D {\bf 39}, 1084 (1989).

\bibitem{Blau}
S.~K.~Blau, E.~I.~Guendelman and A.~H.~Guth,
Phys.\ Rev.\ D {\bf 35}, 1747 (1987).

\bibitem{Garriga:1999yh}
J.~Garriga and T.~Tanaka,
Phys.\ Rev.\ Lett.\  {\bf 84}, 2778 (2000)
[arXiv:hep-th/9911055];
S.~Kanno and J.~Soda,
Phys.\ Rev.\ D {\bf 66}, 083506 (2002)
[arXiv:hep-th/0207029].

\bibitem{soda}
D.~Langlois,
Phys.\ Rev.\ Lett.\  {\bf 86}, 2212 (2001)
[arXiv:hep-th/0010063];
H.~Kodama, A.~Ishibashi and O.~Seto,
Phys.\ Rev.\ D {\bf 62}, 064022 (2000)
[arXiv:hep-th/0004160];
C.~van de Bruck, M.~Dorca, R.~H.~Brandenberger and A.~Lukas,
Phys.\ Rev.\ D {\bf 62}, 123515 (2000)
[arXiv:hep-th/0005032];
K.~Koyama and J.~Soda,
Phys.\ Rev.\ D {\bf 62}, 123502 (2000)
[arXiv:hep-th/0005239];
D.~Langlois, R.~Maartens, M.~Sasaki and D.~Wands,
Phys.\ Rev.\ D {\bf 63}, 084009 (2001)
[arXiv:hep-th/0012044];
K.~Koyama and J.~Soda,
Phys.\ Rev.\ D {\bf 65}, 023514 (2002)
[arXiv:hep-th/0108003].
\bibitem{GL}
R.~Gregory and R.~Laflamme,
Phys.\ Rev.\ Lett.\  {\bf 70}, 2837 (1993)
[arXiv:hep-th/9301052];
S.~Kanno and J.~Soda,
Class.\ Quant.\ Grav.\  {\bf 21}, 1915 (2004)
[arXiv:gr-qc/0311074].
\bibitem{higher}
I.~Navarro and J.~Santiago,
arXiv:hep-th/0402204.



\end{thebibliography}

\end{document}